\pdfoutput=1
\documentclass{JINST}
\usepackage{abhep, abhepexpt, hepnames}

\title{The upgrade of the LHCb trigger system}

\author{Conor Fitzpatrick$^a$, on behalf of the LHCb trigger.\\
  \llap{$^a$}European Organisation for Nuclear Research\\
  Geneva, Switzerland\\
  E-mail: \email{conor.fitzpatrick@cern.ch}}

\abstract{The \LHCb experiment will operate at a luminosity of $2\times10^{33}~\lumiunits$ during LHC Run~3. At this rate the present readout and hardware Level-0 trigger become a limitation, especially for fully hadronic final states. In order to maintain a high signal efficiency the upgraded \LHCb detector will deploy two novel concepts: a triggerless readout and a full software trigger.}


\begin{document}

\section{Introduction}

The \LHCb detector is a precision beauty and charm physics experiment fully instrumented on $2<\eta<5$ at the Large Hadron Collider. The trigger system of the \LHCb experiment during the 2011-2012 Run~1 data taking period consisted of a hardware Level-0 (L0) fixed latency near-detector trigger which reduced the visible bunch crossing rate to $\sim1~\MHz$ at which the detector could be read out, and a flexible software Higher-Level trigger (HLT) applying a range of more advanced selections which further reduced the rate to $\sim5~\kHz$ for offline storage and reprocessing. This configuration allowed \LHCb to produce the largest charm meson samples at high purity~\cite{Aaij:2013wda} and to observe rare \PB decays at high efficiency~\cite{Aaij:2012nna} in a very challenging environment. One novel feature of the Run~1 trigger was that of event deferral: 20\% of all L0 accepted events were stored on disk to be processed by the HLT during the LHC inter-fill time. This made available an effective 25\% more time for processing in the HLT, allowing tracking \pT thresholds to be reduced; a major advantage for charm analyses.  The performance of the Run~1 trigger is described in detail in refs.~\cite{Aaij:2012me,Albrecht:2013fba}. 
\begin{figure}\centering
	\includegraphics[width=0.5\textwidth]{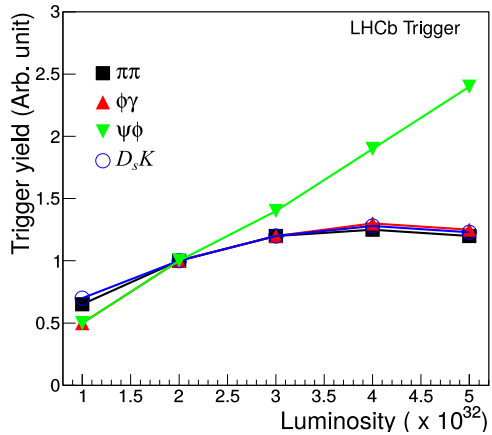}
	\caption{\label{fig:lltrates}Low-level trigger efficiency normalised to that of Run~1 as a function of luminosity for selected hadronic decays. Several modes saturate before the nominal Run~3 luminosity of $2\times10^{33}~\lumiunits$.}
\end{figure}
The Run~2 detector will be mostly unchanged with respect to Run~1, while the trigger software will undergo an incremental update in preparation for Run~3: The event deferral as used in Run~1 will be moved to after the first stage of the HLT where the reduced rate further enhances the time available to process events in the second HLT stage.
In Run~3 the upgraded \LHCb experiment will operate at a factor of five times the Run~1 luminosty. At this luminosity, the $1~\MHz$ readout becomes a bottleneck, as the limited information available to the L0 trigger leads to an unacceptable loss of efficiency, particularly for hadronic final states as indicated in fig.~\ref{fig:lltrates}. As part of the upgrade all of the front-end devices will be replaced in order to read the detector out at the $40~\MHz$ bunch crossing rate. By removing the readout limitation, the L0 trigger can then be moved to the Event Builder (EB) farm as a software algorithm. This software implementation of the former L0, now the LLT, has a timing budget of 2~ms, and can act as a handbrake during the commissioning stages of the upgrade, but is planned to be phased out as the farm size increases~\cite{LHCb:1701361}. The initial LLT algorithm will operate in an identical fashion to the current L0 trigger, but due to the software nature of the LLT, added flexibility allows for modified thresholds and clustering requirements if needed. 

\section{The anatomy of an LHCb upgrade event}
\begin{figure}\centering
	\includegraphics[width=0.5\textwidth]{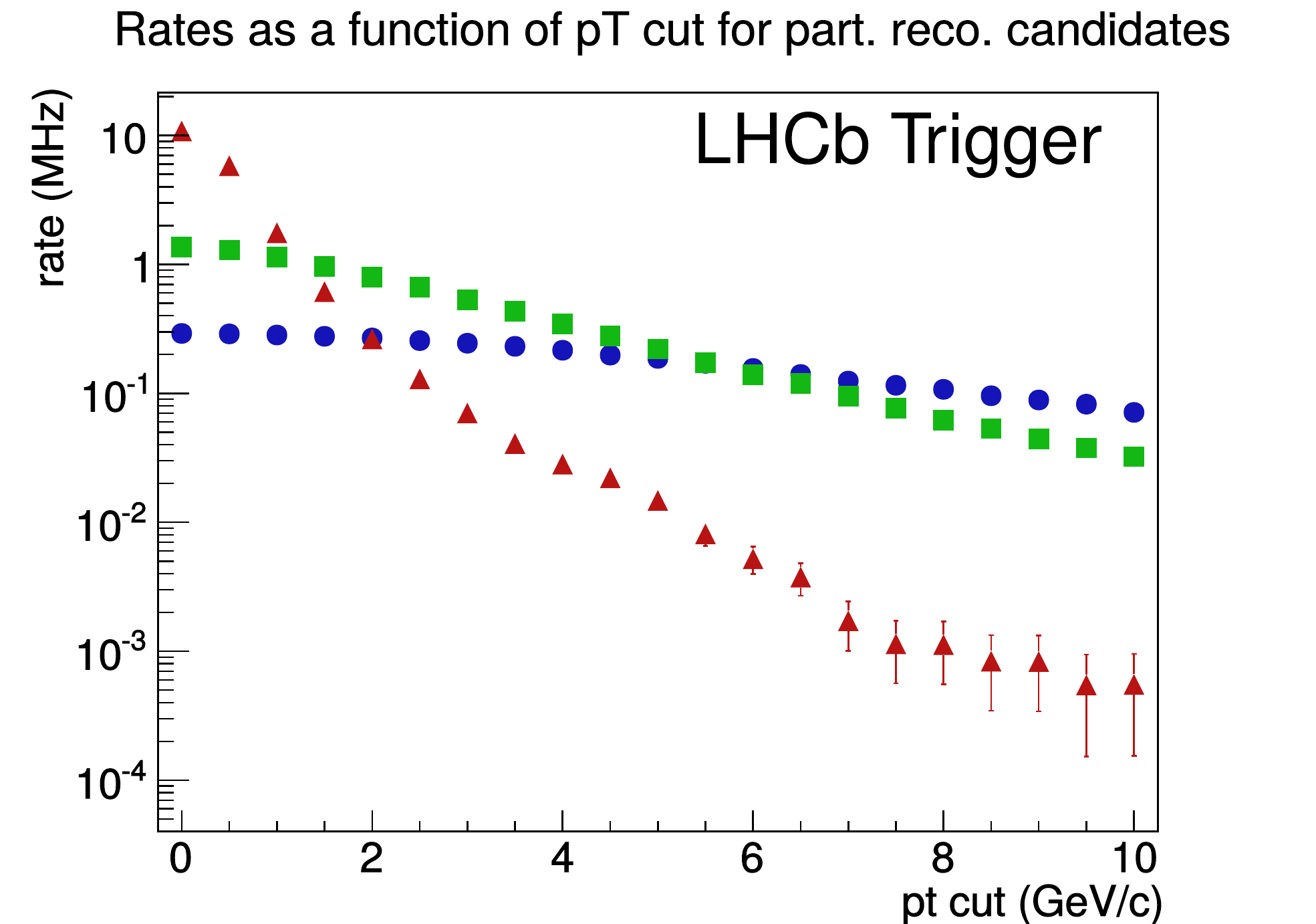}\includegraphics[width=0.5\textwidth]{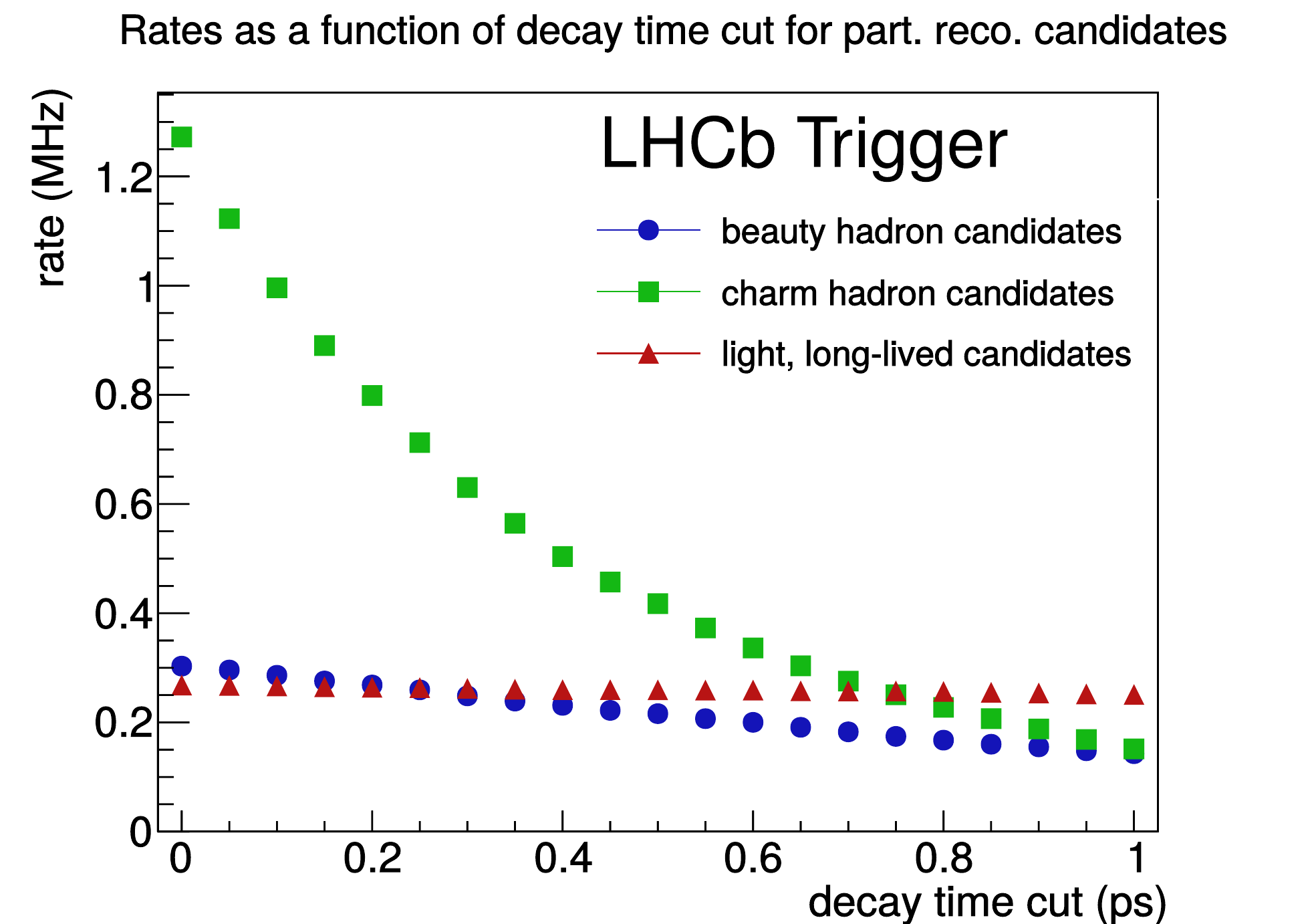} 
	\caption{\label{fig:anatomyrates}Rates in $\MHz$ of events containing reconstructible beauty, charm, and light hadrons passing \pT (left) and decay time (right) requirements in simulated upgrade conditions.}
\end{figure} 
The LHC will provide LHCb with an instantaneous luminosity of $2\times10^{33}~\lumiunits$ with $30~\MHz$ of visible \pp interactions. A study has been performed in which the rates of decays of interest to \LHCb selected by an idealised trigger are presented~\cite{Fitzpatrick:1670985}. At $13~\TeV$ and with a pileup of $\mu=5.2$ almost every event will contain partially reconstructible signal of interest to the \LHCb physics programme. The expected size of an \LHCb event in Run~3 is 100~Kb. Assuming a 100\% efficient trigger, the output rate to offline storage after applying moderate \pT and vertex displacement requirements is 27 GB/s for beauty hadrons, 80 GB/s for charm and 26 GB/s for light, long-lived particles such as \PKshort and \PLambda. The rates as a function of \pT and decay time cut are shown in fig.~\ref{fig:anatomyrates}.  This represents a fundamental change in trigger requirements of \LHCb: It is no longer the case that the trigger must reject background from signal, it must now categorise signal according to physics requirements. Such a strategy requires significantly more information in the trigger than can be provided by low-latency, hardware based solutions. 

\section{The full software trigger}
With the removal of the L0 hardware trigger, \LHCb will be the first hadron collider experiment to deploy a trigger exclusively in software, using off-the-shelf hardware~\cite{LHCb:1701361}. The Run~1 LHCb HLT consisted of software algorithms running on the Event Filter Farm (EFF), $\sim29,000$ logical CPU cores situated in the cavern close to the detector. The upgraded \LHCb trigger will use an expanded version of this farm to select and categorise events at the full collision rate. The advantages of such a system are clear: Software is easily modified, allowing an unprecedented flexibility as and when the \LHCb physics programme changes. It also has the advantage that computing power is readily upgradeable and benefits from the ability to purchase more CPU power for the same price at a later stage. The deferral technique used in Run~1 will be leveraged in Run~2 and Run~3: Subdetector alignment and calibration will be performed while events are buffered to disk, permitting the use of offline-quality Particle ID~\cite{Collaboration:1624074}, and reducing the need for reprocessing of data at a later stage.

\subsection{Tracking at 30~\MHz}
In order to classify events based on their signal content, the upgraded trigger must be strongly aligned to the offline selection requirements. In Run~1, the trigger sequence required that at least one track had either corresponding hits in the muon system or $\pT>1.6~\MeV$. Only after this requirement were selections made. In the Run~3 trigger, \textbf{all} tracks will be made available at the earliest stages of the trigger- the full tracking sequence will be performed upfront~\cite{Head:1670987}. The offline track reconstruction sequence in \LHCb reconstructs tracks close to the interaction point using the Vertex Locator (VELO). Velo tracks are then combined with hits in the tracking stations downstream of the magnet to make forward tracks. 
The offline $\pT$ threshold for forward tracks is $70~\MeV$. All forward tracks that pass this requirement are then available to the remaining tracking algorithms, where information from upstream tracker which sits between the magnet and VELO are added, and PV finding, kalman track fit and particle ID are performed. 
In the trigger tracking sequence the same algorithms are used, but the sequencing and configuration are modified to reconstruct the most valuable tracks first so that slower, specialised track reconstruction algorithms only need to be made later in the decision making process. The upstream tracker is used before forward tracking in order to reduce the rate of tracks projected into the downstream tracking stations. For this we will require a loose \pT cut of $200~\MeV$ on upstream tracks. Forward tracks are then required to have $\pT>500~\MeV$ prior to the PV finding algorithm. In this scenario all tracks which pass each stage are kept, only the order in which the algorithms are executed is changed. The total timing for this tracking sequence is estimated to be $5.4$ ms, while the timing budget of the event filter farm is estimated to be $13$ ms. The upgrade tracking sequence is made possible by the advanced design of the upgraded tracking system~\cite{Collaboration:1624070,Collaboration:1647400}.  

\subsection{The inclusive beauty trigger}

\begin{figure}\centering
	\includegraphics[width=0.5\textwidth]{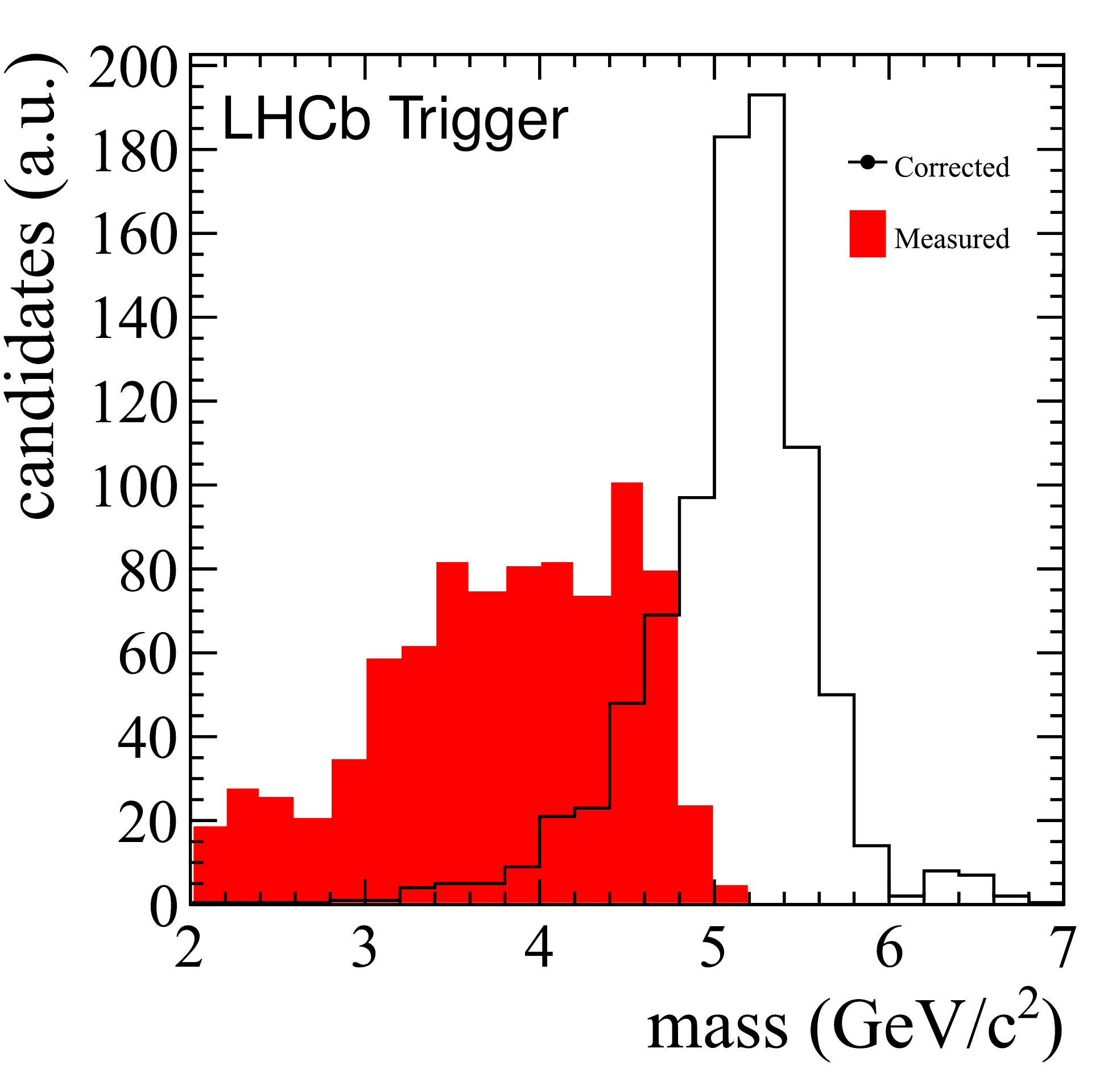}
	\caption{\label{fig:mcorr}Simulated \HepProcess{\PB\to\PKstar\Pmu\Pmu} events. The reconstructed
		2-body mass is shown in red and the corrected mass (see text for
	definition) is shown in black.}
\end{figure}
\begin{figure}
	\includegraphics[width=0.33\textwidth]{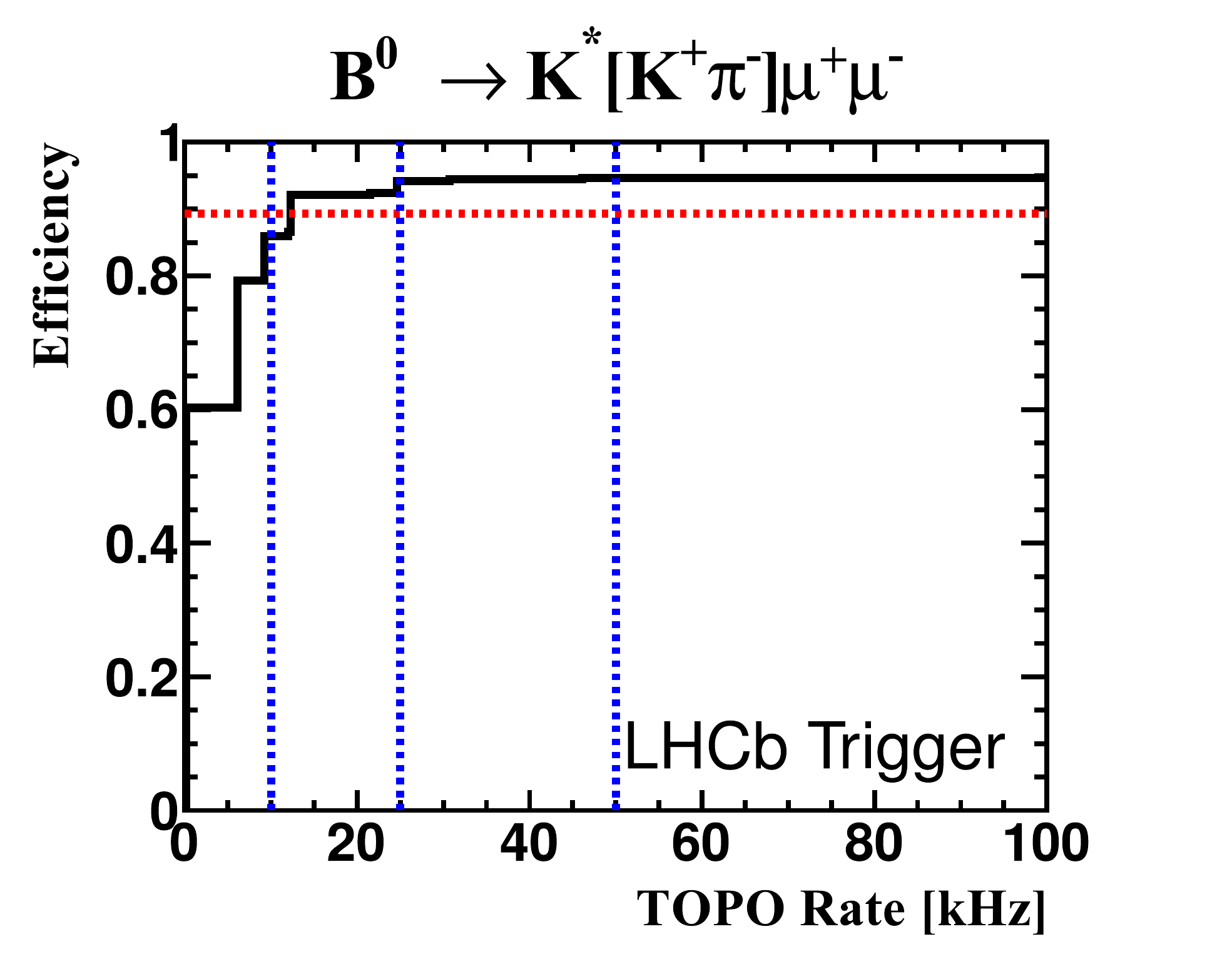} 
	\includegraphics[width=0.33\textwidth]{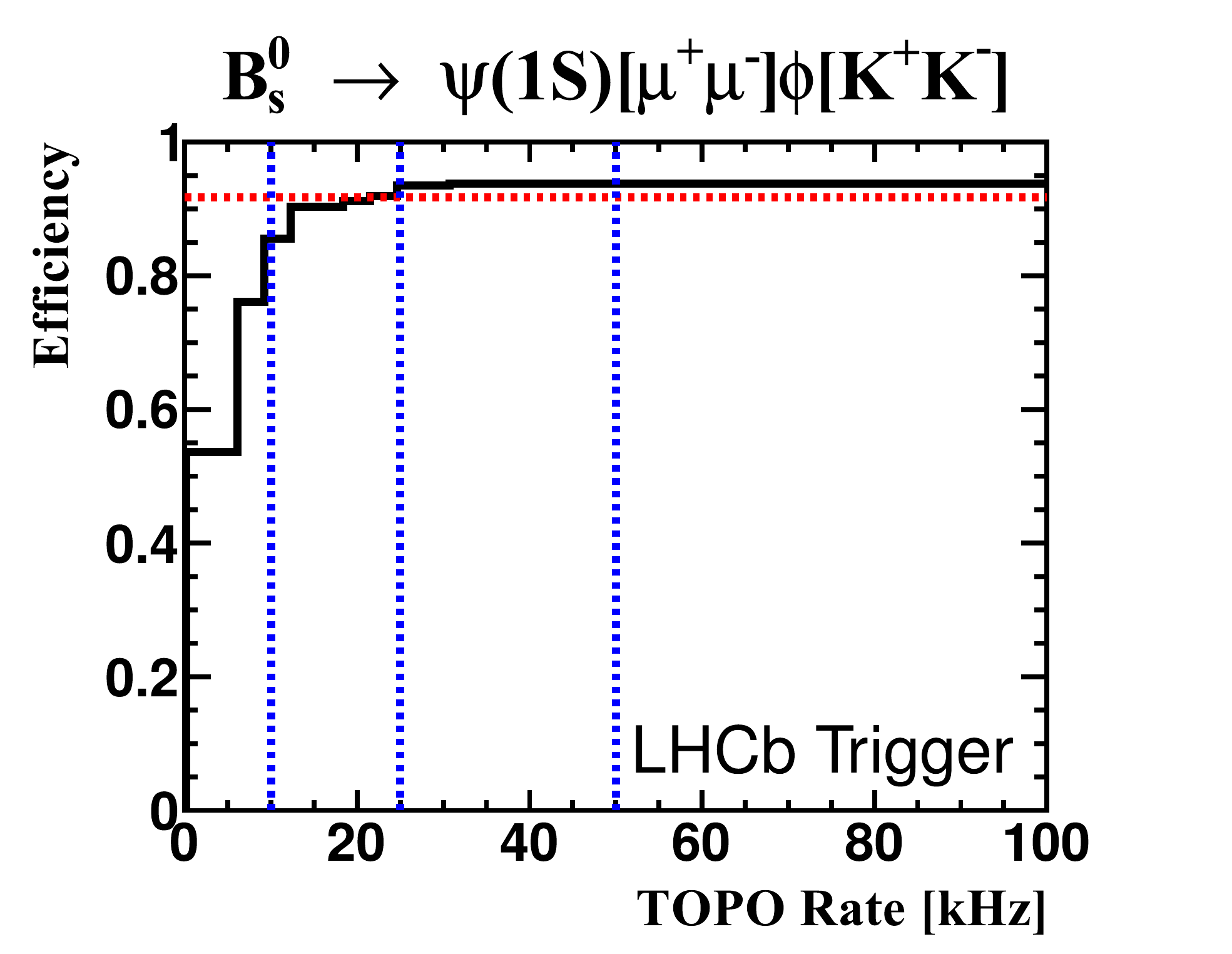}
	\includegraphics[width=0.33\textwidth]{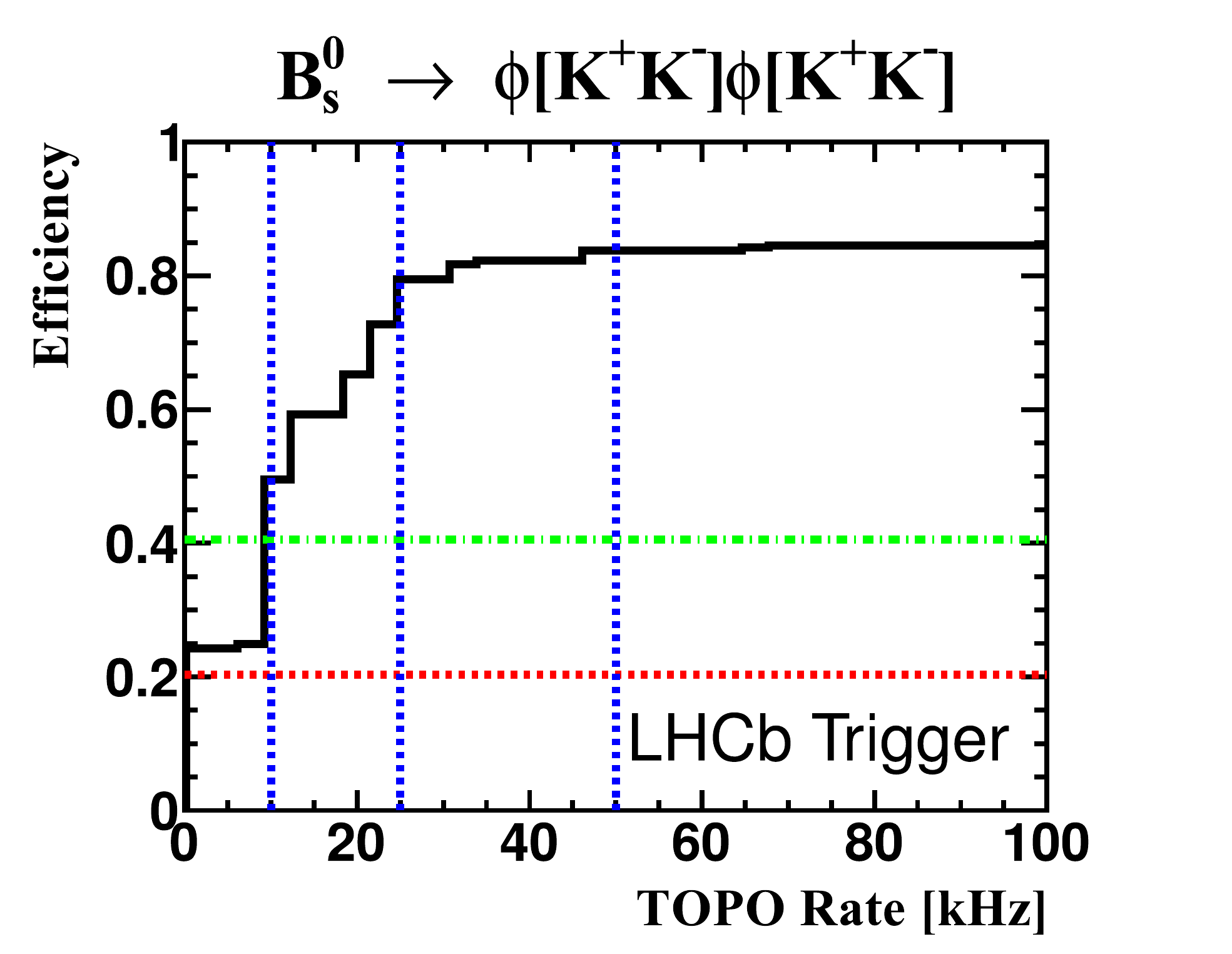} \\
	\includegraphics[width=0.33\textwidth]{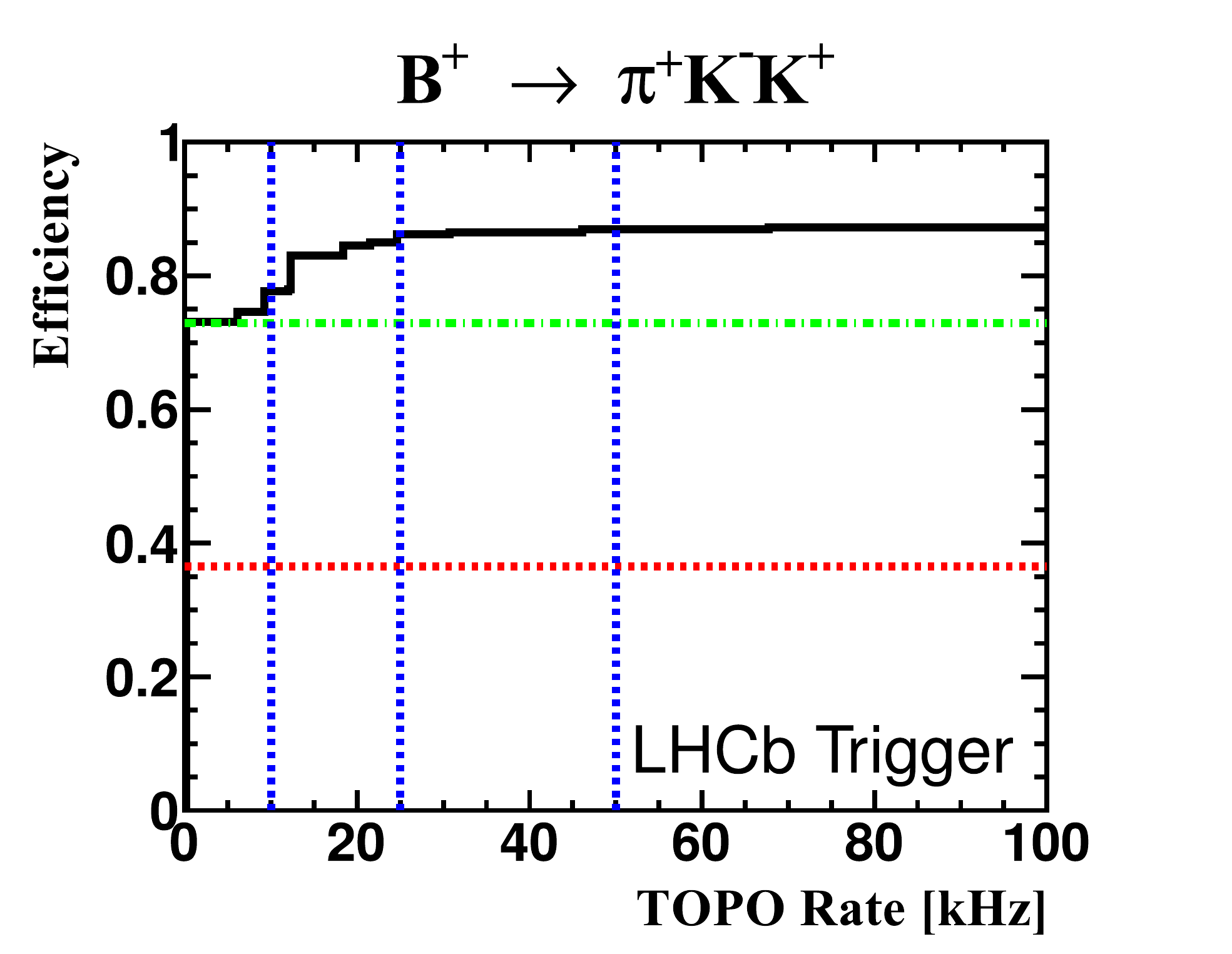} 
	\includegraphics[width=0.33\textwidth]{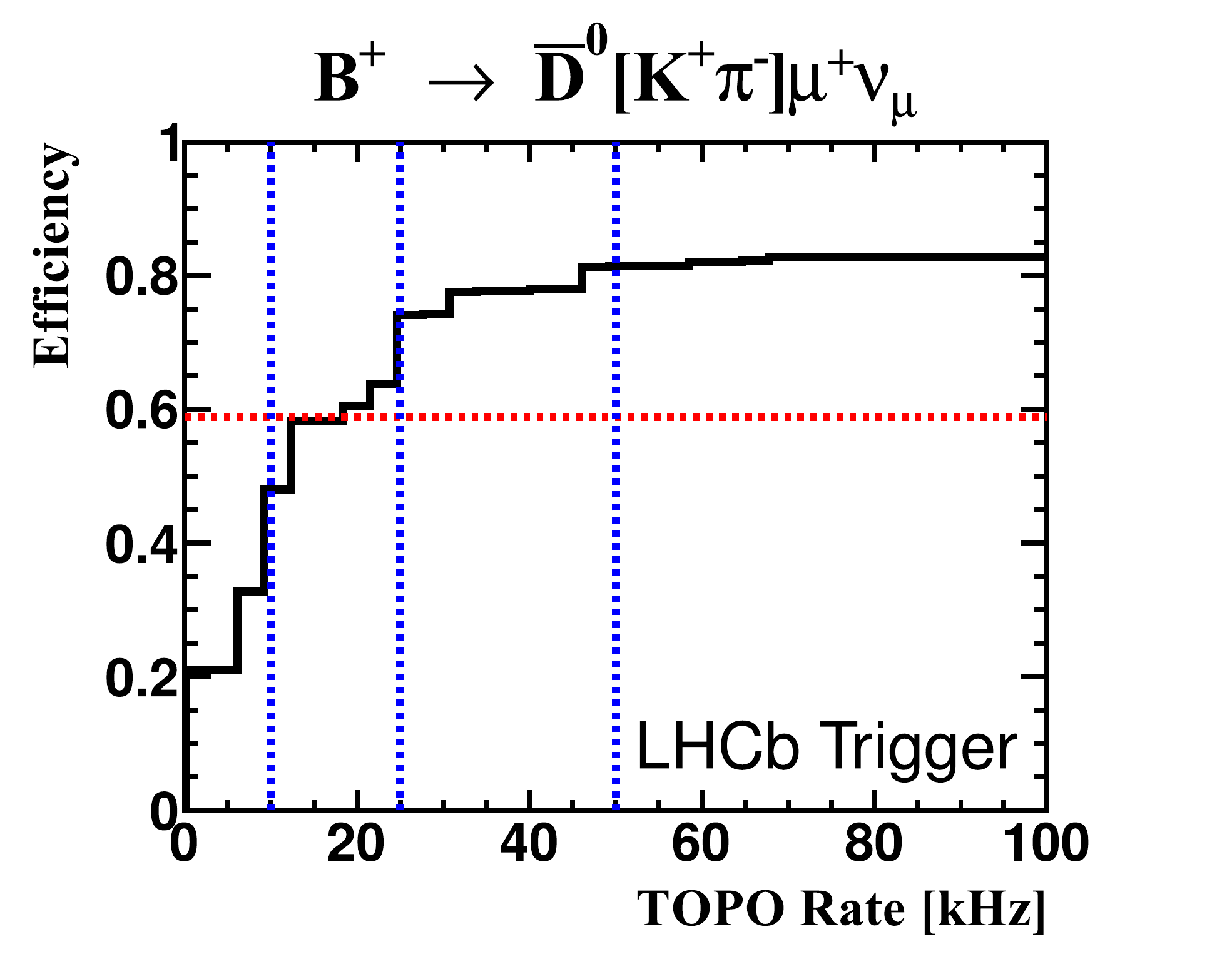}
	\includegraphics[width=0.33\textwidth]{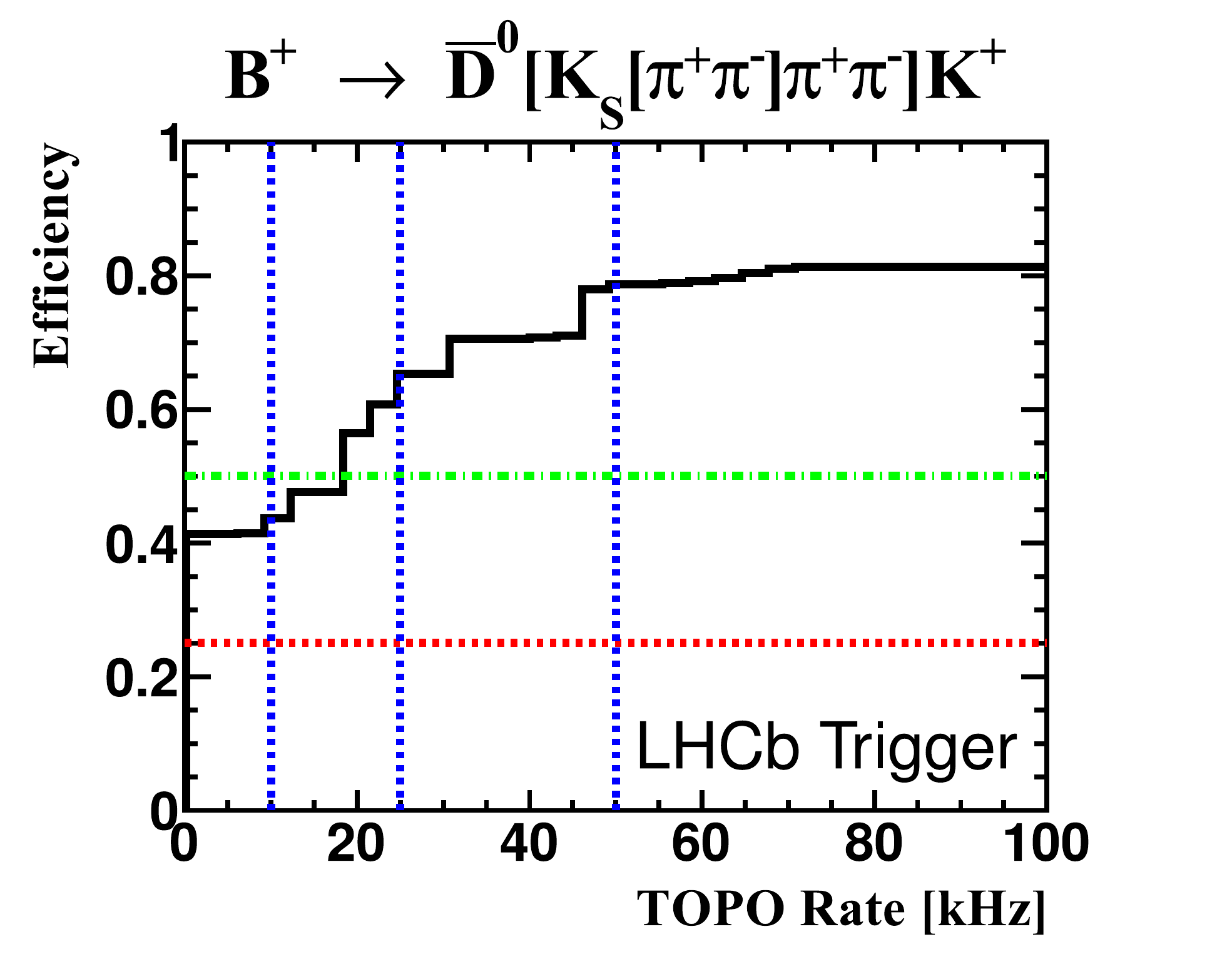}
	\caption{\label{fig:toporates}Efficiencies as a function of topo output rate for several benchmark decay modes. Red lines indicate the efficiency during Run~1. Green lines indicate twice the Run~1 efficiency. Blue lines indicate three prospective output rates at 10, 25 and $50~\kHz$ respectively.}
\end{figure}
After the tracking sequence, the remaining timing budget is available to apply offline-quality trigger selections. In Run~1 we demonstrated that complex, multivariate trigger selections were possible in the LHCb data taking environment: the majority of analyses using $b$-hadrons at \LHCb made use of the topological n-body trigger~\cite{Gligorov:1384380}. The topological n-body trigger is an inclusive trigger which combines successive 2,3 and 4-body track combinations.  A novel BDT multivariate selection~\cite{2013JInst...8P2013G} is used. The BDT makes use of several inputs, consisting of impact parameter and flight distance significances as well as \pT and $m_{\text{corr}}$. The corrected mass, $m_{\text{corr}}$, is particularly powerful. This is defined as: 
\begin{equation}
	m_{\text{corr}}=\sqrt{m^{2} + |\pT_{\text{miss}}|^{2}} + |\pT_{\text{miss}}|,
\end{equation}
where $\pT_{\text{miss}}$ is the missing momentum transverse to the direction of flight of the candidate assuming it originates from its best primary vertex. This variable is particularly powerful in selecting partially reconstructed \PB decays as it permits selection based on a proxy to the fully reconstructed candidate mass. Fig. ~\ref{fig:mcorr} shows the invariant mass of partially reconstructed 2-body candidates for the decay \HepProcess{\PB\to\PKstar\Pmu\Pmu}, and their corrected mass. 

The topological trigger in Run~1 selected 99.9\% pure \bbbar signal with a modest output rate. For Run~3 the removal of the readout limitation leads to appreciable efficiency gains after retraining of the BDT. Fig.~\ref{fig:toporates} indicates the efficiency with respect to offline selectable candidates for a range of benchmark modes as a function of the output rate. The total timing of the Topo n-body trigger is $0.1$~ms, meaning that a large fraction of the LHCb physics programme can be triggered in a negligible fraction of the available timing budget.  

\subsection{Lifetime unbiased triggers}
\begin{figure}\centering
	\includegraphics[width=0.5\textwidth]{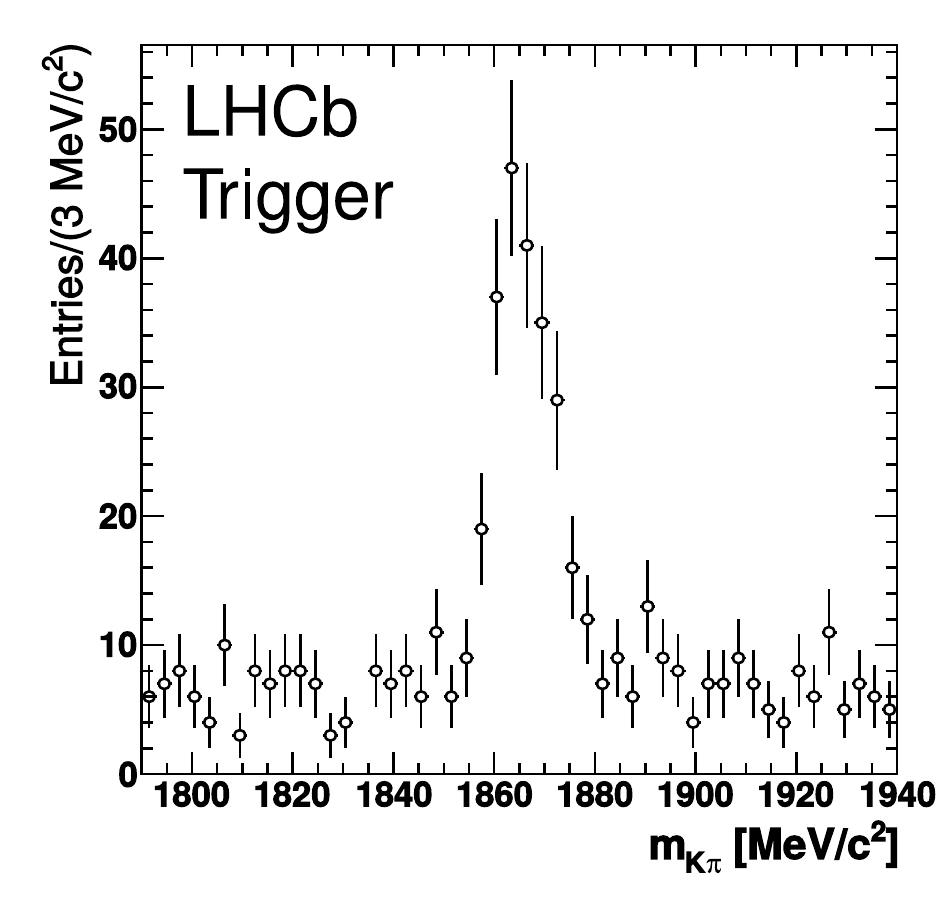}
	\caption{\label{fig:charmpeak}Invariant mass distribution of \HepProcess{\PD\to\PK\Ppi} candidates in simulated upgrade events selected using the lifetime unbiased charm trigger. The data sample corresponds to approximately 30 ms of data taking.}
\end{figure}

The ability to perform offline-like selections at trigger level means that lifetime unbiased triggering is possible for the first time in a hadron collider environment. For several analyses at 
LHCb the modelling of decay time acceptances are or will become dominant systematic uncertainties. In general these arise from the use of impact parameter or flight distance significance requirements to suppress prompt backgrounds, as are used to good effect in the topological trigger. For time dependent analyses it is better to apply a direct cut on the decay time such that all events above the cut are free of acceptance effects. This is made possible in an all-software trigger where the decay time resolution is comparable to that determined offline. For \PB decays efficiencies of 90\% with respect to offline selected events are possible above a $0.3~\ps$ decay time requirement with negligible rate and timing. For charm modes similar timings are possible but rates remain high even at high purity due to the large charm cross section: Fig~\ref{fig:charmpeak} indicates the trigger-selected \HepProcess{\PD\to\PK\Ppi} yield and purity to be expected within as little as 30 ms of Run~3 startup. Suppression of Cabibbo-favored \PDstar-tagged \PD decays are however. Cabibbo-suppressed decays are two orders of magnitude less common, and can be kept while the Cabibbo-favored mode can be prescaled to reasonable rates. Discrimination between the  Cabibbo-suppressed and Cabibbo-favored modes is achieved through the use of offline-quality particle ID information. 

\section{Conclusions}

The upgraded LHCb trigger represents a turning point in the design of hadron collider trigger systems. By adopting an all-software approach running on commercially available hardware an inexpensive, flexible and scalable design is possible. The upgraded trigger will perform offline-quality tracking at the full collision rate in less than half of the available timing budget. This leaves room for a number of inclusive and exclusive trigger selections. The use of event buffering for alignment and calibration permits selections that would otherwise only be available in the offline analysis environment. Subject to output bandwidth requirements, gains of up to a factor of four in efficiency are possible for several LHCb benchmark modes.  

\bibliographystyle{unsrt}
\bibliography{wit2014_ConorFitzpatrick.bib}
\end{document}